\documentclass[conference]{IEEEtran}
\IEEEoverridecommandlockouts

\usepackage[caption=false,font=footnotesize]{subfig}

\usepackage[T1]{fontenc}
\usepackage[utf8]{inputenc}
\usepackage{csquotes,tabularx,uulm-logos,tabu,makecell} 
\usepackage{xspace,hyperref}
\usepackage[ruled]{algorithm2e}

\usepackage{enumitem}
\setlist{noitemsep}

\usepackage[textsize=footnotesize,disable]{todonotes}
\usepackage{tikz}
\usepackage{amsmath}
\usepackage{multirow}
\usepackage{tikz-layers}

\usetikzlibrary{positioning, fit, calc, shapes, arrows, fadings}

\tikzfading[name=glass,
inner color=transparent!100,
outer color=transparent!90]

\usepackage[backend=biber,sortcites=true,minnames=5,maxnames=10,mincitenames=1,maxcitenames=2,bibencoding=utf8,
isbn=false,giveninits=true,doi=false,
natbib]{biblatex}
\AtEveryBibitem{%
	\clearlist{language}%
    \clearfield{doi}%
    \clearfield{pages}%
	\ifentrytype{article}{%
		\clearfield{isbn}%
	}{}%
	\ifentrytype{techreport}{%
	}{%
		\clearfield{urlyear}%
	} %
	\ifentrytype{inproceedings}{%
		\clearfield{pages}%
		\clearlist{location}%
		\clearfield{month}%
        \clearlist{publisher}%
        \iffieldundef{series}{}{\clearfield{booktitle}}%
	}{%
	}%
	\ifentrytype{misc}{%
	}{%
		\ifentrytype{online}{%
		}{%
			\clearfield{url}%
			\clearfield{urldate}%
		}
	}
}
\def\BibTeX{{\rm B\kern-.05em{\sc i\kern-.025em b}\kern-.08em
    T\kern-.1667em\lower.7ex\hbox{E}\kern-.125emX}}
\bibliography{PRE}
\bibliography{Tools}

\newcommand{\refappendix}[1]{\hyperref[#1]{Appendix~\ref*{#1}}}

\newcommand{\eg}{, e.\,g.,\xspace}
\newcommand{\ie}{, i.\,e.,\xspace}

\newcommand{\ecdfraw}{\ensuremath{\widehat E}}
\newcommand{\ecdf}{\ensuremath{\ecdfraw_k}\xspace}
\newcommand{\ecdfzero}{\ecdf} 
\newcommand{\Dmatrix}{\ensuremath{\mathbf{D}}}
\newcommand{\Dset}{\ensuremath{\mathrm{D}}}

\newcommand{\slink}[2]{\ensuremath{s_{{#1},{#2}}^\mathrm{link}}}
\newcommand{\dlink}{\ensuremath{d_{i,j}^\mathrm{link}}}
\newcommand{\density}{\rho}
\newcommand{\shortl}{s_l}
\newcommand{\minmedians}{\operatorname{minmed}}

\newcommand{\mean}[1]{\ensuremath{\overline{#1}_\mathrm{arithm}}}
\newcommand{\median}[1]{\ensuremath{\overline{#1}_\mathrm{median}}}
\newcommand{\std}{\ensuremath{\sigma}}
\newcommand{\round}{\ensuremath{\operatorname{round}}}
\DeclareMathOperator*{\argmax}{\arg\!\max}
\DeclareMathOperator*{\argmin}{\arg\!\min}

\newcommand{\tp}{\text{TP}}
\newcommand{\fp}{\text{FP}}
\newcommand{\tn}{\text{TN}}
\newcommand{\fn}{\text{FN}}

\newcommand{\fscore}{\ensuremath{\mathrm{F}_\frac{1}{4}}}
\newcommand{\fbeta}{\ensuremath{\mathrm{F}_\beta}}

\begin{document}

\title{Network Message Field Type Clustering for Reverse Engineering of Unknown Binary Protocols}

\author{
	 \IEEEauthorblockN{Stephan Kleber and Frank Kargl}
	 \IEEEauthorblockA{\textit{Institute of Distributed Systems},\\ \textit{Ulm University}, Germany\\
	 	\{stephan.kleber,frank.kargl\}@uni-ulm.de}
	 \and
	 \IEEEauthorblockN{Milan Stute and Matthias Hollick}
	 \IEEEauthorblockA{\textit{Secure Mobile Networking Lab},\\ \textit{Technical University of Darmstadt}, Germany\\
	 	\{mstute,mhollick\}@seemoo.de}
}

%


\maketitle

\IEEEpubid{\begin{minipage}{\textwidth}\ \\[24pt]
        \copyright 2020 IEEE. Personal use of this material is permitted. Permission from IEEE must be
        obtained for all other uses, in any current or future media, including
        reprinting/republishing this material for advertising or promotional purposes, creating new
        collective works, for resale or redistribution to servers or lists, or reuse of any copyrighted
        component of this work in other works.\\
        IEEE DSN Workshop on Data-Centric Dependability and Security 2022.
        DOI: \href{https://doi.org/10.1109/DSN-W54100.2022.00023}{10.1109/DSN-W54100.2022.00023}
\end{minipage}} 

\IEEEpubidadjcol
\begin{abstract}
Reverse engineering of unknown network protocols based on recorded traffic traces enables security analyses and debugging of undocumented network services.
One important step in protocol reverse engineering is to determine data types of message fields.
Existing approaches for binary protocols 
(1) lack comprehensive methods to interpret message content and determine the data types of discovered segments in a message
and 
(2) assume the availability of context, which prevents the analysis of complex and lower-layer protocols. 
Overcoming these limitations, we propose the first \emph{generic} method to analyze message field data types in unknown binary protocols
by clustering of segments with the same data type.
Our extensive evaluation shows that our method in most cases provides clustering of up to 100\,\% precision at reasonable recall.
Particularly relevant for use in fuzzing and misbehavior detection, we increase the coverage of message bytes over the state-of-the-art to 87\,\% by almost a factor of 30.
We provide an open-source implementation to allow follow-up works.
\end{abstract}


\pdfinfo{info}

\section{Introduction}

Protocol reverse engineering (PRE) based on traffic traces aims to infer the specification of unknown network protocols by analyzing traces of network messages typically gained from observing communication of devices implementing this protocol.
PRE is often applied to understand malware communication and uncover data exfiltration~\cite{cho_inference_2010}, to configure smart fuzzers~\cite{gascon_pulsar:_2015}, or to validate the correct and secure design and implementation of undocumented network services~\cite{wen_protocol_2017}.
As a recent example, PRE was necessary to discover a severe vulnerability in the proprietary Apple Wireless Direct Link (AWDL) protocol stack~\cite{stute_one_2018}, enabling a zero-click exploit~\cite{beer_google_2020} affecting all of Apple's iOS-based product lines.
Thus, PRE helps in identifying security implications that result from the intended or unintended use of a specific unknown protocol.


PRE based on traffic traces encompasses the uncovering of message types, message formats, semantics, and behavior of the protocol.
During this kind of analysis, semantic deduction is one of the most tedious and scarcely automated tasks~\cite{stute_one_2018,kroll_aristoteles_2021}.
One step in semantic analysis is inferring the data type or value domain of fields which can help, for example, to more efficiently configure smart fuzzers or to identify exfiltration~\cite{cho_inference_2010,bermudez_towards_2016,wressnegger_zoe:_2018}.  
While some methods are available that recognize single field data types and correlations of values,
no approach determines relations between fields by their value similarity that can be used to interpret the message contents.

\textbf{Contribution.}
This paper proposes a novel method to automatically cluster field data types.
We base this inference on the analysis of segments\ie subsequences of network messages.
We propose to distinguish segments into \emph{clusters of the same field data type} according to their similarity to each other without actually identifying the data type.
The resulting knowledge of segments with identical type simplifies follow-up analyses as value domains can be inferred and spoofing or fuzzing require this knowledge.
%
As opposed to previous approaches~\cite{bossert_towards_2014,bermudez_towards_2016,cui_discoverer:_2007}
and particularly important for security assessments of custom and proprietary protocols, we make very few assumptions about the format and sequence of messages.
We summarize our main contributions as follows:
\begin{itemize}
	\item We design the first method to cluster field types of network messages and do so without a limiting set of individual rules per type, making our approach applicable to a wide range of protocols with diverse and unanticipated data representations.
    \item Based on empirical observations of typical network protocols, we devise a fully-automated parameter selection method that is use-case-specific to the clustering of field values.
	\item We implement our method as well as FieldHunter~\cite{bermudez_towards_2016} and CSP~\cite{goo_protocol_2019} and make all three publicly available.\footnote{%
        \url{https://github.com/vs-uulm/nemesys},
        \href{https://github.com/vs-uulm/fieldhunter}{fieldhunter}, and
        \href{https://github.com/vs-uulm/goo-csp}{goo-csp}
        } 
	\item Through extensive evaluation of both well-known and proprietary protocols, we show that our method 
	on average achieves an F-score of 0.92 for field type clustering.
        At the same time, coverage of 87\,\% message bytes exceeds the state-of-the-art by almost a factor of 30. 
\end{itemize}

\IEEEpubidadjcol
\section{Related Work}
\label{sec:related}

\begin{figure*}
	\centering
	\begin{tikzpicture}[
every node/.style={
	outer sep=.5ex, inner sep=2ex,
	font=\sffamily\footnotesize, align=left}
]
\coordinate (top) at (0, 2em);
\coordinate (bot) at (0,-2em);

\node[label={above:\autoref{sec:preprocessing}}] at (0,0)                 (ppr) {Preprocess\\ trace};
\node[label={above:\autoref{sec:segmentation}}, right=1.5em of ppr]       (seg) {Heuristic\\ segments};  
\node[label={above:\autoref{sec:dissimilarity}}, right=1em of seg]        (dis) {Calculate\\ dissimilarity};
\node[label={above:\autoref{sec:auto-configuration}}, right=1em of dis]   (aut) {Clustering parameter\\ auto-configuration};
\node[label={above:\autoref{sec:clustering}}, right=1em of aut]           (clu) {Cluster segments\\ on dissimilarities};
\node[label={above:\autoref{sec:cluster-refinement}}, right=1.5em of clu] (fin) {Refine\\ clusters};

\begin{scope}[on background layer, 
every path/.style={transform canvas={xshift={.5ex}}, fill=uulm-akzent!60}
]
\foreach \x in {ppr} {
	\path
	(\x.west |- top) to (\x.west |- bot)
	to ([xshift=-1em]\x.east |- bot) to ([xshift=+1em]\x.east)
	to ([xshift=-1em]\x.east |- top) to cycle;
}
\foreach \x in {seg,dis,aut,clu} {
	\path
	([xshift=-1.5em]\x.west |- top) to ([xshift=+.5em]\x.west) to
	([xshift=-1.5em]\x.west |- bot)
	to ([xshift=-1.5em]\x.east |- bot) to ([xshift=+.5em]\x.east)
	to ([xshift=-1.5em]\x.east |- top) to cycle;
}
\foreach \x in {fin} {
	\path
	([xshift=-2em]\x.west |- top) to (\x.west)
	to ([xshift=-2em]\x.west |- bot)
	to (\x.east |- bot)
	to (\x.east |- top) to cycle;
}
\end{scope}

\begin{scope}[
every node/.style={align=left, anchor=west, font=\sffamily\footnotesize},
goa/.style={->, line width=2pt, transform canvas={xshift=2ex}, uulm-akzent},
rotup/.style={rotate=10, anchor=south west},
rotdn/.style={rotate=-10, anchor=north west}
]
\coordinate (goa) at (0, -2em);
\node[rotdn] (seggoa) at (goa -| seg.west) {Netzob | CSP\\ | NEMESYS};
\node[rotdn] (disgoa) at (goa -| dis.west) {Canberra};
\node[rotdn] (fingoa) at (goa -| aut.west) {\texttt{min\_samples} and $\varepsilon$};
\node[rotdn] (clugoa) at (goa -| clu.west) {DBSCAN};

\end{scope}
\end{tikzpicture}
	\vspace{-1em}
	\caption{Clustering of common kinds of message content data.}
	\label{fig:mode3}
\end{figure*}
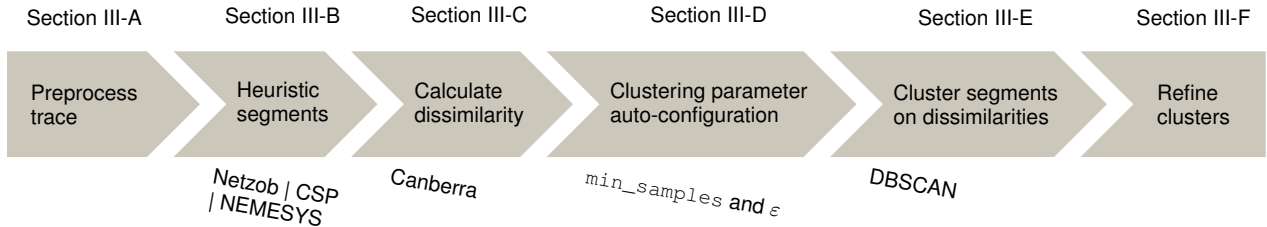

Surveys have proposed to structure the overall PRE process into multiple phases~\cite{duchene_state_2018, kleber_survey_2019}.
Typical phases are data collection into traces, feature extraction, message type identification, message format inference, semantic deduction, and behavior model reconstruction.
Existing PRE approaches differ substantially for textual and binary protocols, where analysis of textual protocols is often considered the easier task~\cite{cui_discoverer:_2007, bermudez_towards_2016, kleber_survey_2019, duchene_state_2018}.
Thus, we focus on providing a solution for binary network protocols.
Previous work, like Discoverer~\cite{cui_discoverer:_2007}, PRISMA~\cite{krueger_learning_2012}, Netzob~\cite{bossert_towards_2014}, \citeauthor{goo_protocol_2019}~\cite{goo_protocol_2019}, NEMETYL~\cite{kleber_message_2020}, and many others,
focused either on the message type and format or on the behavior model of unknown protocols.
While most phases are well covered in literature, approaches specifically addressing the interpretation of the message contents\ie semantic deduction of fields, are rare.
All existing methods are rule-based~\cite{cui_discoverer:_2007, bossert_towards_2014, bermudez_towards_2016, goo_protocol_2019}\ie they consist of a finite set of individual heuristic rules that explicitly deduce the semantics of a predefined small number of single specific field types, like number, identifier, or network address.
%
%
We consider FieldHunter~\cite{bermudez_towards_2016} the state-of-the-art approach as it has also been re-applied in recent work~\cite{goo_protocol_2019}.
If either, the protocol uses a representation of data types that was not anticipated in any of the heuristic rules, or the encapsulation is unknown so that context like addresses is not available, FieldHunter fails to work.
As there is no public implementation, we re-implemented FieldHunter and evaluate its results in comparison to our approach in this paper.

As opposed to FieldHunter and all previous rudimentary field type inference approaches mentioned in this section, we aim for a more generic goal than using only a limited number of individual heuristics for field types:
Our work is the first to propose clustering of arbitrary data types of message \emph{fields}.
We stress that we do not attempt to identify particular field types so that we are not limited to a predefined set of data types.
In this work, we do not consider clustering whole messages into different message types since previous work like \eg Discoverer~\cite{cui_discoverer:_2007}, PRISMA~\cite{krueger_learning_2012}, Netzob~\cite{bossert_towards_2014}, or NEMETYL~\cite{kleber_message_2020}, already achieve this goal.

Our approach relies on message segmentation and different methods might achieve different quality.
To consider segmentation accuracy in our evaluation, we compare three existing segmenters that work with variable-length fields:
Netzob~\cite{bossert_towards_2014} is based on sequence alignment,
CSP~\cite{goo_protocol_2019} applies frequency analysis of byte-strings,
and NEMESYS~\cite{kleber_nemesys:_2018} uses statistical properties of the message contents
to discern one approximated field candidate from the other, forming heuristic segmentations of unknown binary messages.
%
Furthermore, to compare segments to each other, we use the Canberra dissimilarity~\cite{kleber_message_2020}, which we originally proposed for message type identification.
We now apply it directly for clustering of segments while, in contrast, its original usage was to be input for sequence alignment of messages.
%

\section{Clustering Data Types}
\label{sec:approach}
\label{sec:step1}


Our approach provides the means to cluster independent message segments into what we call \emph{pseudo data types} without any further knowledge about the protocol.
Individual steps of this process are outlined in \autoref{fig:mode3}.
It is a heuristical method to cluster the same types of data into groups of similar field contents.
Having such clusters of segments throughout different messages of a trace reveals relationships of values between messages regardless of the byte positions of the segments within each message.
We call the resulting clusters \emph{pseudo} data types because, at this point, we do not know which data type or field semantic the cluster represents.
An analyst can still use this knowledge as basis to analyze the properties of the clustered segments and infer their semantic meaning.
We now discuss the individual steps.

\subsection{Preprocessing}\label{sec:preprocessing}
We first \textbf{preprocess} each raw trace.
This step includes filtering for the desired protocol and de-duplicating payloads.
Our analysis method exploits variances in the contents of messages, so duplicates carry no additional information.

\subsection{Segmentation}\label{sec:segmentation}
We define a \textbf{field} in a binary protocol specification as a sequence of bytes at a specific position in a message, with a specific data type such as an integer, a sequence of chars, or a timestamp, and a value domain.
In contrast, we define a \textbf{segment} to be a field candidate determined from the inference that---in an optimal case---matches the true field from the unknown protocol specification.
Segmentation is an important prerequisite for \emph{characterizing the contents of messages}, which is needed to determine the segments' data types, infer their semantics, and ultimately deduce an accurate field definition.

To obtain \textbf{segments} from the messages in the traces, we split individual messages into subsequences.
Messages of known protocols can be segmented by using dissectors, like those provided by Wireshark.\footnote{\url{https://www.wireshark.org}} 
While dissectors are unavailable for unknown protocols to reliably determine message fields, heuristic approximations can be used to find probable field boundaries and obtain segments that are field candidates.
Thus, we require a segmenter that can identify segments in unknown protocols.
Available solutions include 
Netzob~\cite{bossert_towards_2014}, \citeauthor{goo_protocol_2019}~\cite{goo_protocol_2019}, and NEMESYS~\cite{kleber_nemesys:_2018}.
We evaluate these three segmenters that have the advantage that they work equally well for protocols of fixed structure and such with dynamic field lengths differing between messages.
The idenfified segments are now treated as candidates for protocols fields.


\subsection{Dissimilarity}\label{sec:dissimilarity}
To calculate a similarity measure for segments, we interpret each of these as a \textbf{vector} of byte values.
We then calculate a normalized \textbf{dissimilarity} value for each pair of segments using the so-called Canberra dissimilarity~\cite{kleber_message_2020}, which extends the better-known Canberra distance~\cite{lance_computer_1966} to vectors of different dimensions.
We store the pairwise dissimilarities between all segments in a dissimilarity matrix $\Dmatrix$.

We exclude segments from the analysis that are only one byte long as coincidental similarity of arbitrary single bytes throughout messages prevents meaningful analysis of such short segments.
Using alternative analysis methods, like frequency analysis, these one-byte segments can later be reincorporated in the analysis.
Furthermore, we consider duplicate segment values only once since they increase the computational load without adding new information for the subsequent clustering.

The dissimilarity values for each pair of remaining unique segments serve as affinity values to guide clustering by Density-Based Spatial Clustering of Applications with Noise (DBSCAN)~\cite{ester_density-based_1996} in the next steps.

\subsection{Auto-Configuration}\label{sec:auto-configuration}
Before clustering, we need to configure two parameters of DBSCAN:
the minimum number of elements to form a density core \texttt{min\_samples} and a measure $\varepsilon$ of the least density to be considered part of a cluster.
Normally, these parameters need to be configured and tuned manually.
For unsupervised and fully automated, configuration-less clustering, we present a new method to automatically determine the parameters for DBSCAN from the properties of the segments identified in the previous step.

The $\mathbf{\varepsilon}$ \textbf{auto-configuration} searches for the knee point in the empirical cumulative distribution function (ECDF)~\cite{van_der_vaart_empirical_1998} $\ecdf(d)$ of the dissimilarities between the $k$-nearest-neighbors ($k$-NN) of unique segments.
For each trace, a set of functions $\ecdf$ exists, one ECDF for each $k$.
An ECDF is an evenly-spaced step function, jumping by $\frac{1}{n}$ for each of the $n$ samples with a measured value $d$.
In our case, the samples are the segments $s_i$ and $s_j$ in a trace, and their measured value is the dissimilarity $d(s_i,s_j)$.
Applied to the $k$-NN function, the ECDF's value thereby is the fraction of all segments in a trace that have a Canberra dissimilarity less or equal to their respective $k$th nearest-neighbor.
The ECDF plots the changes in distances between neighbors.
A clear drop $d_\kappa$, located at the knee point $\ecdf(\kappa)$, is then considered a suitable choice for $\varepsilon$ that allows DBSCAN to reliably detect cluster boundaries.

\begin{algorithm}[b!]
	\caption{$\varepsilon$ auto-configuration}
	\label{alg:epsilon}

   \SetKwInOut{Input}{input}\SetKwInOut{Output}{output} 
    
    \SetKwProg{Fn}{function}{}{end}
    \SetKwFunction{kNN}{kNN}
    \SetKwFunction{ecdFn}{ecdf}
    \SetKwFunction{bSpline}{bSpline}
    \SetKwFunction{Kneedle}{Kneedle}

    \Input{
        Set of dissimilarities $\Dset$\;
        Sensitivity parameter of Kneedle $S$\;  
        Smoothness parameter of B-Spline interpolation $s$\;  
    }
    \Output{$\varepsilon$}
    \BlankLine
	
    \Fn{\kNN{$\Dset$, $k$}}{
        \emph{Determine the $k$-NN of all segments represented in} $\Dset$\;
        \Return Dissimilarities of all segments' $k$th-NN\;
    }
    \BlankLine

    \ForEach{$2 \leq k \leq \round(\ln n)$}{
        $\ecdf \leftarrow \ecdFn{\kNN{\Dset, k}}$\;
        $\widehat{B}_k \leftarrow \bSpline{\ecdf, s}$\;
    } 
	
	$k' \leftarrow \argmax\limits_k\  \delta\widehat{B}_k$ 
    \tcc*{Value of the maximum increase in distance}
	
	$d_\kappa \leftarrow$ \Kneedle{$\widehat{B}_{k'}, s$}\;
	$\varepsilon \leftarrow d_\kappa$\;
\end{algorithm}

Of all possible $\ecdf$, we want to dynamically select $k$ in such a way that its ECDF has the most distinct drop in the density of the segment similarity. 
The function that contains the most distinct change in distances between neighbors has the sharpest knee point.
Consequently, we search for the $\ecdf$ with the sharpest knee, with sharpness measured as the value of the $\delta d$ at the maximum of $\delta \ecdf$.
\autoref{alg:epsilon} describes the process to select the desired $k$.
We iterate $k$ only between $2$ and $\round(\ln n)$ to limit the number of unnecessary calculations.
This is sufficient since the sharpest relevant knees always are in the distance distributions of neighboring segments.

To determine the rightmost knee point in $\ecdfzero$ with the selected $k$, we apply the Kneedle algorithm~\cite{satopaa_finding_2011}.
Kneedle requires smoothing of the ECDF, for which we use a spline, to remove local statistical fluctuations before accepting it as input.
\autoref{fig:knee} illustrates the ECDF, the $\max (\delta \ecdf)$, the effect of smoothing, and the detected knee, used as $\varepsilon$ with segments generated from a trace of 1,000 NTP messages.
For the 2nd parameter \texttt{min\_samples}, we note that DBSCAN is not very sensitive and setting it to $\ln n$ simply prevents scattering large traces into too many small clusters.

\subsection{Clustering}\label{sec:clustering}
Next, we \textbf{cluster segments} with DBSCAN using the determined parameters $\varepsilon$ and \texttt{min\_samples}.
DBSCAN is a popular and efficient clustering algorithm that makes no assumptions about the shape of clusters, does not require the target number of clusters as input, and treats outliers as noise.
These properties set it apart from traditional clustering methods\eg $k$-means or spectral clustering, which are unsuitable for our purpose since we do not know the shape and number of clusters.
For other clustering methods, 
\todo{omit?}
like agglomerative clustering, affinity propagation, or support vector machines, 
automating the tuning of the parameters for previously unseen traces is challenging.
In comparison, DBSCAN's main advantage is that we can design a method to directly derive its parameters from the dissimilarity distribution of a trace, as described in the previous section.
Thus, the proposed configuration procedure is completely automated and requires no re-training or iterative tuning for new traces as may be the case for other clusterers.

The algorithm identifies high-density cores within noisy data and determines them to be clusters of similar segments.
The segment density is high in areas where segments have a low dissimilarity to each other.
Each cluster groups similar segments and thus comprises fields of a common data type.
We validate the underlying assumption that clusters regularly coincide with data types in the first part of our evaluation (\autoref{sec:eval-validate}).
%

\todo{omit paragraph?}
In rare cases where the dissimilarity distribution leads to multiple knees in the ECDF, the so-determined $\varepsilon$ does not denote suitable densities to cluster field data types and is too large.
In this situation, a single large cluster contains more than 60\,\% of the segments that are not considered noise.
To prevent this and instead select the next smaller knee for an $\varepsilon$, we consider only a subset of the original $\ecdf$.
More specifically, we repeat the whole $\varepsilon$ auto-configuration process for a $\ecdf'$ that is trimmed to the last detected knee $\kappa$, which becomes the rightmost value.
Thus, $\ecdf' = \ecdf(\{d < d_\kappa : d \in \Dset \})$.
We then cluster with the new $\varepsilon$ value.

\newcommand{\nodi}{.7em}
\todo{algorithm2e?!}
\begin{figure}
	\begin{tikzpicture}[
			every node/.style={font=\sffamily\scriptsize},
			leg/.style={anchor=west, align=left, node distance=1em}
	]
		\node[inner sep=0] (img) {%
			\includegraphics[width=.93\linewidth]{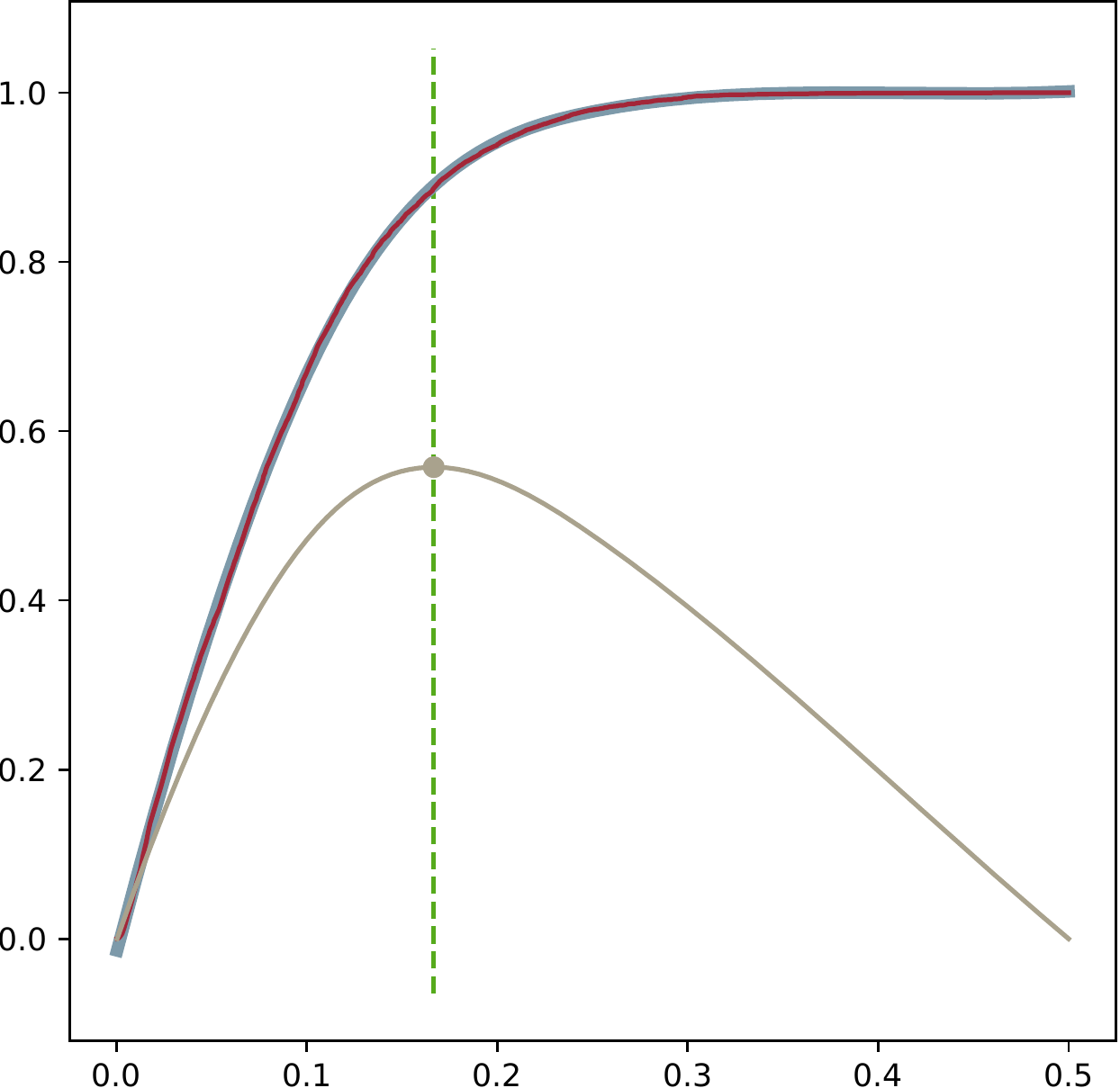}};
		\node[anchor=north] (xaxis) at (img.south) {dissimilarity};
		\node[anchor=south, rotate=90] (yaxis) at (img.west) {fraction of neighboring segments};
		
		\begin{scope}[shift={(-2.2,1.5)}]
			\clip (0,0) circle (3em+1pt);
			\draw[line width=2pt, uulm-akzent, fill=white] (0,0) circle (3em);
			\node[inner sep=0] (base) at (1.1,-0.1) {\includegraphics[width=16em, trim=98px 230px 220px 90px, clip]{graphics/knee-ntp_SMIA-20111010_maxdiff-1000}};
			\draw[line width=2pt, black!60] (0,0) circle (3em);
			\clip (0,0) circle (3em-1pt);
			\fill[fill=uulm, path fading=glass] (0,0) circle (8em);
		\end{scope}
		
		\coordinate (legend) at ({$(img.west)!.6!(img.east)$}|-{$(img.north)!.72!(img.south)$});
		
		\begin{scope}[on above layer]
			\node[leg] (smoothed) at (legend) {smoothed ECDF $\ecdfraw_2$};
			\node[below=\nodi of smoothed.base west, leg] (raw) {raw ECDF $\ecdfraw_2$};
			\node[below=\nodi of raw.base west, leg] (diff) {differences of ECDF $\ecdfraw_2$};
			\node[below=\nodi of diff.base west, leg] (knee) {knee = 0.167};
			
			\coordinate (symL) at ([xshift=-3.5ex]legend);
			\coordinate (symR) at ([xshift=-.3ex]legend);

			\draw[uulm, line width=3pt] (symL|-smoothed) to (symR|-smoothed);
			\draw[uulm-in, line width=1pt] (symL|-raw) to (symR|-raw);
			\draw[uulm-akzent, line width=1pt] (symL|-diff) to (symR|-diff);
			\draw[uulm-mawi, line width=1pt, dashed] (symL|-knee) to (symR|-knee);
		\end{scope}
		\node[fit=(symL)(smoothed)(diff), fill=white, opacity=.7] (legendbox) {};

		\node[anchor=west] (maxdiff) at ({$(img.west)!.46!(img.east)$}|-{$(img.north)!.4!(img.south)$}) {maximum difference $\max (\delta \ecdfraw_2)$};
		\draw[->, bend right, line width=.7pt] (maxdiff.west) to +(-14pt,-5pt);
		
		\node (adjeps) at ({$(img.west)!.65!(img.east)$}|-{$(img.north)!.2!(img.south)$}) {adjusted $\varepsilon$ shifted from knee};
		\draw[->, bend left, looseness=.7, line width=.7pt] ({$(img.west)!.4!(img.east)$}|-{$(img.north)!.21!(img.south)$}) to +(10pt,7pt);
	\end{tikzpicture}
	\caption{ECDF $\ecdfraw_2$ and its knee, detected by Kneedle, at a dissimilarity of 0.167 used as $\varepsilon$.}
	\label{fig:knee}
\end{figure}

\subsection{Cluster Refinement}\label{sec:cluster-refinement}
In situations where the variability of the field values in the input trace is not uniformly distributed, multiple clusters may result for the same data type. 
This overclassification is not only a limitation of DBSCAN and we noticed that similar alternatives\eg HDBSCAN and OPTICS, suffer from the same effect.
We favor DBSCAN since it provides more possibilities to fine-tune the cluster boundaries for our use case.
During pilot analyses of known protocols,
we observed that overclassified clusters are often linked via sparsely populated but detectable areas.
To mitigate the overclassification, we introduce an additional step: \textbf{cluster refinement}.

In cluster refinement, we aim to automate the detection and merging of clusters that are nearby and have a similar density. 
For any two clusters $c_i$ and $c_j$, we define the \emph{link segment} $\slink{i}{j}$ as the segment in $c_i$ that is closest to $c_j$\ie
$$
\slink{i}{j} = \argmin_{s_i \in c_i} d(s_i,s_j), \quad \forall s_j \in c_j
$$
$\dlink$ is the distance between the link segments $\slink{i}{j}$ and $\slink{j}{i}$.

Using this definition, we propose two heuristic cluster merge conditions: 
(1) The clusters are very close-by, and the densities within an $\varepsilon$ around the link segments are similar.
(2) The clusters are somewhat close-by, and the whole clusters have similar densities.
We quantify closeness and density differently in both conditions
since we intend to capture different notions of closeness and density, i.e., local $\varepsilon$-density at elements linking clusters and density of clusters as a whole.

In \textbf{Condition 1}, \emph{clusters are very close by} if the link-dissimilarity is less than the mean $\mean{D(c_i)}$ of the set of pairwise dissimilarities in $c_i$ or respectively $c_j$:
$$\dlink < \max(\mean{\Dset(c_i)}, \mean{\Dset(c_j)})$$
%
Further, we use a density definition for an $\varepsilon$-neighborhood around the nearest points between similar clusters.
W.\,l.\,o.\,g., we define $\shortl = \slink{i}{j}$.
The \emph{density $\rho$ within an $\varepsilon$ around the link segment} in $c_i$ with the set of dissimilarities $\Dset(c_i)$, 
$$
\Dset(\varepsilon,{\shortl}) = \{ d(\shortl, s_c) \  : \ d(\shortl, s_c) \leq \varepsilon, \  \shortl \neq s_c, \  s_c \in c_i \},
$$
with $\Dset(\varepsilon,{\shortl}) \subseteq \Dset(c_i)$, is thus defined by
$
\density_{\varepsilon}(\shortl) = \median{ \Dset(\varepsilon,{\shortl}) }.
$
We observed that a suitable $\varepsilon$ is half of the maximum extent $d_\mathrm{max}$ of the cluster with the fewer segments:
$\varepsilon = \frac{d_\mathrm{max}}{2}$.
The density around the link segment in $c_j$ is defined accordingly.
Finally, $\varepsilon$-densities around link segments are considered similar if their difference is less than
\emph{$\varepsilon\density$Threshold}:
$$
\left| \density_\varepsilon(\slink{i}{j}) - \density_\varepsilon(\slink{j}{i}) \right| < \textit{$\varepsilon\density$Threshold}
$$

\textbf{Condition 2} allows a larger cluster distance but has a stronger density requirement.
Close-by here means closer than the mean between both cluster's \enquote{neighbor densities} normalized to the extent of the cluster.
To formalize this, we need $\minmedians(i)$, the median values for the 1st-nearest neighbors that are the minimum distances for each segment to any other in $c_i$, respectively $c_j$:
$$
\minmedians(i) = \median{\{ \min_{s_b \in c_i}( \{ d(s_a, s_b) : s_a \neq s_b \} ) : s_a \in c_i \}}
$$
In terms of Condition 2, \emph{clusters are somewhat-close-by} if
$$
\dlink < \cfrac{\cfrac{\minmedians(i)}{\mean{\Dset(c_i)}} + \cfrac{\minmedians(j)}{\mean{\Dset(c_j)}}}{2}
$$
As expression of the overall density in the cluster, we use $\minmedians$.
In contrast to the $\varepsilon$ density defined above, whole \emph{clusters have similar density} if
$$
\left| \minmedians(i) - \minmedians(j) \right| < \textit{neighborDensityThreshold}.
$$

The selection of the values
$\textit{$\varepsilon\density$Threshold} = 0.01$ and 
$\textit{neighborDensityThreshold} = 0.002$
results from empirical observation of real-world protocols.


\medskip
Unlike overclassification, occasional underclassification combines different field data types in one cluster.
This may be the case if a single value is similar to a group of others but has a distinct function, like an enumeration value.
To compensate for this, we split clusters if they have extremely polarized value occurrences\eg they exhibit many unique values, together with very few, very high occurring ones.
For this purpose, occurrences are defined as the count $\left| b \right|$ of segments $s \in c$ with a value $b$.
We count all different values $b_i$ and calculate the standard deviation 
$\std(\{ \left| b_i \right| : 0 \leq i < \left| c' \right| \})$ 
for this cluster $c'$, with $b_i$ being the value of one or multiple segments $s$. 
To interpret the counts of values of one cluster, we use the percent rank $\operatorname{PR}$~\cite{roscoe_fundamental_1975} as a combined measure of the value occurrence frequency and value diversity.
A $\operatorname{PR}(c',F) = 95$ means that 95\,\% of the value counts in cluster $c'$ are below the given occurrence frequency of interest $F$.
We select $F$ depending on the cluster size to be $\ln \left| c' \right|$.
Thus, the same value is the pivot to split the cluster into two subclusters containing all segments with value count $\left| b_i \right| \leq F$ and another for $\left| b_i \right| > F$ for $i$ enumerating all distinct values of segments in the cluster.
Consequently, if $\operatorname{PR}(c',F) > 95 \  \land \  \std(\{ \left| b_i \right| : 0 \leq i < \left| c' \right| \}) > F$, we split $c'$ at the pivot $F = \ln \left| c' \right|$.

\subsection{Summary}
After completing this fully automated procedure, we now have generated vectors for each segment, calculated their pairwise Canberra dissimilarity, determined a suitable $\varepsilon$ value, clustered the segments using DBSCAN, and refined the clusters.
This completes the clustering of segments into pseudo data types. 
Next, our evaluation will show how accurately this can be done.

\section{Evaluation}
\label{sec:evaluation}

Using our proof-of-concept implementation, we evaluate two different aspects of our approach.
First, we validate that data types can be clustered accurately by segment similarity.
Second, we evaluate the accuracy achievable by using heuristic segmentation in the absence of ground truth.
We illustrate the validity of these two aspects by 
clustering statistics.

\subsection{Metrics and Setup}

For a quantitative representation of the clustering quality, we calculate \textbf{precision} $P$ and \textbf{recall} $R$ of the clusters compared to the true data types
by the number of true positives ($\tp$), false positives ($\fp$), and false negatives ($\fn$)
as:
$$P = \frac{\tp}{\tp + \fp} \text{\quad and \quad} R = \frac{\tp}{\tp + \fn}$$

For clustering into more than two clusters, $\tp$, $\fp$, true negatives ($\tn$), and $\fn$ are defined combinatorically via the correct and incorrect pairwise assignments of unique segments, as described by \citeauthor{manning_introduction_2009}~\cite{manning_introduction_2009}.
Hence, the number of positives and negatives for $m$ clusters $c_i$ are given as:
$$\tp + \fp = \sum_{i} \binom{ \left| c_i \right| }{2} \qquad \text{and} \qquad \tn + \fn = \sum_{i, j} \left( \left| c_i \right| \cdot \left| c_j \right| \right),$$
where $j = \lbrace 0 \dots (m-1) \rbrace \setminus i$.
The true positives are:
$$\tp = \sum_{i} \sum_{l} \binom{\left| t_{i,l} \right|}{2},$$
where $t_{i,l}$ denotes the segments of data type $l$ in cluster $i$.
The false negatives are defined through the missed true pairs by false assignments to different clusters and to the noise.
Thus, the count of false negatives is given by the sum of both kinds of false negatives through:
\pagebreak[0]
\begin{align*}
\fn &= \sum_{i} \sum_{l} \frac{( \left| t_l \right| - \left| t_{i,l} \right| ) \cdot \left| t_{i,l} \right| }{2} \\
&+ \sum_{l} \binom{\left| t_{n,l} \right|}{2} 
+ \sum_{l} \frac{( \left| t_l \right| - \left| t_{n,l} \right| ) \cdot \left| t_{n,l} \right| }{2},
\end{align*}
where $t_{n,l}$ are the segments of data type $l$ assigned to the noise.

To compare the quality between different protocols and input segments, we require an overall quality measure.
Therefore, we calculate the $\fscore$ \textbf{score} from precision and recall.
The $\fbeta$ score is a common measure for the clustering accuracy and defined by the harmonic mean of precision and recall~\cite{van_rijsbergen_information_1979}.
Parameter $\beta$ adjusts the weight of precision and recall in the mean.
With $\beta = \frac{1}{4}$, we place four times more emphasis on precision than recall.
We decided on this weighting since precise clusters are crucial for a meaningful data type analysis in protocols. 
At the same time, low recall diminishes the coverage but does not reduce the validity of the overall analysis result.
As coverage we define the ratio between the number of inferred bytes and all bytes of all messages in a trace.
Since coverage refers to the number of bytes and precision and recall to segment pairs the result statistics are not directly correlated.

\newcommand{\ravel}[1]{\multicolumn{1}{|[el]r|[el]}{#1}} 
\newtabulinestyle {el=0.75pt uulm-akzent} 

\begin{table*}%
	\begin{minipage}[t]{0.34\linewidth}%
		\scriptsize\sffamily\raggedright
		\caption{Clustering statistics for data type clustering from ground truth.}
		\begin{tabular}{@{}lr|rr|rrr@{}}
			&&&&&& \\
			\textbf{proto.} & \textbf{msg.s} & \textbf{fields} & \multicolumn{1}{c|}{$\mathbf{\varepsilon}$} & \multicolumn{1}{c}{$P$} & \multicolumn{1}{c}{$R$} & $\fscore$ \\ \hline
			
			DHCP & 1000 & 1017 & 0.172 & 0.96 & 0.93 & 0.95 \\
			DNS  & 1000 & 839  & 0.063 & 1.00 & 0.95 & 1.00 \\
			NBNS & 1000 & 734  & 0.049 & 1.00 & 0.91 & 0.99 \\
			NTP  & 1000 & 3822 & 0.121 & 1.00 & 0.96 & 1.00 \\
			SMB  & 1000 & 1175 & 0.218 & \textbf{0.59} & \textbf{0.70} & \textbf{0.60} \\
			AWDL & 768  & 2190 & 0.096 & 1.00 & 0.77 & 0.98 \\\hline
			
			DHCP & 100  & 229  & 0.212 & 0.76 & \textbf{0.66} & \textbf{0.75} \\
			DNS  & 100  & 114  & 0.143 & 1.00 & 0.89 & 0.99 \\
			NBNS & 100  & 131  & 0.121 & 1.00 & 0.56 & 0.96 \\
			NTP  & 100  & 470  & 0.198 & 1.00 & 1.00 & 1.00 \\
			SMB  & 100  & 171  & 0.169 & 0.92 & \textbf{0.48} & \textbf{0.87} \\
			AWDL & 100  & 396  & 0.101 & 0.99 & 0.59 & 0.95 \\
			AU   & 123  & 316  & 0.366 & 1.00 & 0.44 & 0.93
		\end{tabular}
		\label{tab:gt-clusters-comb}
		
		\vspace{1ex}
		\textbf{Worst cases} are printed in bold.
	\end{minipage}
	\hfill
	\begin{minipage}[t]{0.62\linewidth}
		\scriptsize\sffamily\raggedleft
		\caption{Combinatorial clustering statistics and coverage (cov.)\\ for pseudo data types of heuristic segments.}
		\begin{tabular}{@{}||lllr|lllr|lllr@{}}  
		  \multicolumn{4}{||c|}{\textbf{Netzob}} &
		  \multicolumn{4}{c|}{\textbf{NEMESYS}} &
		  \multicolumn{4}{c}{\textbf{CSP}} \\
		 
		  \textbf{$P$} & 
		  \textbf{$R$} &
		  \textbf{$\fscore$} &
          \textbf{cov.} &
		  \textbf{$P$} & 
		  \textbf{$R$} &
		  \textbf{$\fscore$} &
          \textbf{cov.} &
		  \textbf{$P$} & 
		  \textbf{$R$} &
		  \textbf{$\fscore$} &
          \textbf{cov.}  \\ \hline
		  
		  \multicolumn{4}{||c|}{\textbf{fails}} &
		  \textbf{0.88} &
		  \textbf{0.33} &
		  {\color[HTML]{00AE00} \textbf{0.80}} &
          99\,\% &
		  0.85 &
		  0.35 &
		  0.79 &
          99\,\% \\
		  \textbf{0.99} &
		  \textbf{0.96} &
		  {\color[HTML]{00AE00} \textbf{0.99}} &
          100\,\% &
		  1.00 &
		  0.85 &
		  0.99 &
          99\,\% &
		  0.95 &
		  0.76 &
		  0.93 &
          99\,\% \\
		  0.99 &
		  0.74 &
		  0.97 &
          100\,\% &
		  \textbf{1.00} &
		  \textbf{0.95} &
		  {\color[HTML]{00AE00} \textbf{1.00}} &
          100\,\% &
		  0.90 &
		  0.30 &
		  0.80 &
          99\,\% \\
		  \textbf{0.94} &
		  \textbf{0.90} &
		  {\color[HTML]{00AE00} \textbf{0.94}} &
          88\,\% &
		  0.65 &
		  0.61 &
		  0.64 &
          95\,\% &
		  0.68 &
		  0.53 &
		  0.67 &
          73\,\% \\
		  \multicolumn{4}{||c|}{\textbf{fails}} &
		  \textbf{0.57} &
		  \textbf{0.02} &
		  {\color[HTML]{C9211E} \textbf{0.24}} &
          81\,\% &
		  0.38 &
		  0.01 &
		  0.11 &
          79\,\% \\
		  \textbf{1.00} &
		  \textbf{0.93} &
		  {\color[HTML]{00AE00} \textbf{0.99}} &
          99\,\% &
		  0.80 &
		  0.16 &
		  0.64 &
          98\,\% &
		  \multicolumn{4}{c}{\textbf{fails}} \\ \hline

		  0.44 &
		  0.11 &
		  0.38 &
          83\,\% &
		  \textbf{0.83} &
		  \textbf{0.52} &
		  {\color[HTML]{00AE00} \textbf{0.80}} &
          87\,\% &
		  0.24 &
		  0.07 &
		  0.21 &
          87\,\% \\
		  \textbf{0.98} &
		  \textbf{0.86} &
		  {\color[HTML]{00AE00} \textbf{0.97}} &
          100\,\% &
		  0.98 &
		  0.75 &
		  0.96 &
          95\,\% &
		  0.46 &
		  0.13 &
		  0.40 &
          87\,\% \\
		  0.91 &
		  0.85 &
		  0.91 &
          93\,\% &
		  \textbf{0.98} &
		  \textbf{0.56} &
		  {\color[HTML]{00AE00} \textbf{0.94}} &
          99\,\% &
		  0.93 &
		  0.32 &
		  0.84 &
          82\,\% \\
		  \textbf{0.98} &
		  \textbf{0.23} &
		  {\color[HTML]{00AE00} \textbf{0.82}} &
          65\,\% &
		  0.87 &
		  0.01 &
		  0.19 &
          39\,\% &
		  0.71 &
		  0.00 &
		  0.05 &
          65\,\% \\
		  0.59 &
		  0.20 &
		  0.53 &
          81\,\% &
		  \textbf{0.84} &
		  \textbf{0.12} &
		  {\color[HTML]{C9211E} \textbf{0.63}} &
          67\,\% &
		  0.42 &
		  0.11 &
		  0.36 &
          74\,\% \\
		  \textbf{0.99} &
		  \textbf{0.51} &
		  {\color[HTML]{00AE00} \textbf{0.94}} &
          90\,\% &
		  0.59 &
		  0.05 &
		  0.35 &
          92\,\% &
		  0.99 &
		  0.43 &
		  0.92 &
          92\,\% \\
		  \multicolumn{4}{||c|}{\textbf{fails}} &
		  1.00 &
		  0.05 &
		  0.49 &
          84\,\% &
		  \textbf{1.00} &
		  \textbf{0.14} &
		  {\color[HTML]{C9211E} \textbf{0.74}} &
          81\,\%
		\end{tabular}
		\label{tab:segmenters-clusters-comb}
		
		\vspace{0.9ex}
		\textbf{Best ({\color[HTML]{00AE00} green}) and worst ({\color[HTML]{C9211E} red}) cases} are printed in bold and colored.
	\end{minipage}%
\end{table*}

The messages we use for developing our approach are collected from \textbf{traces} of the binary network protocols DHCP, DNS, NBNS, NTP, and SMB.%
\footnote{
	Dynamic Host Configuration Protocol (RFC 2131), 
	Domain Name System (RFC 1035), 
	NetBIOS Name Service (RFC 1002), 
	Network Time Protocol (RFC 958), and
	Server Message Block 
}
All traces are publicly available.%
\footnote{%
	DHCP, NBNS, NTP, and SMB extracted from \url{http://download.netresec.com/pcap/smia-2011/};
	DNS extracted from \url{https://ictf.cs.ucsb.edu/archive/2010/dumps/ictf2010pcap.tar.gz}}
%
In addition, we also use traces of two proprietary protocols, namely Apple Wireless Direct Link (AWDL) and Auto Unlock (AU).
AWDL is a Wi-Fi-based link-layer protocol for peer-to-peer communication.
AU is a proprietary distance bounding protocol.%
\footnote{\url{https://support.apple.com/en-us/HT206995}}
Both protocols were not publicly documented until they recently were reverse engineered manually.
The reverse-engineered specification of AWDL, including a dissector, is publicly available~\cite{stute_one_2018}, and we had access to a private Wireshark dissector of the AU protocol.
Thus, both protocols constitute realistic use cases where ground truth is available to verify our results.
We use only protocols with ground truth to compare our results to, which is not available for truly unknown protocols.
Otherwise, statistical analysis of the quality of our approach would not be possible.

As the source of the ground truth, we parse the Wireshark dissectors' output for each message.
All evaluated protocols are binary, while DNS, DHCP, SMB, and AWDL also contain embedded char sequences.
The binary fields of DNS, NBNS, and NTP have fixed length, while DHCP, SMB, AWDL, and AU use a mix of fixed and variable-length fields.
\todo{probably abbreviate}
DHCP, DNS, NBNS, SMB, AWDL, and AU support varying numbers of fields in different messages while NTP has a fixed structure.
Thus, our set of traces represents a wide variety of different protocol properties.
For the evaluation of clustering and recognition, we truncate the traces to achieve comparable results.
We truncate to 100 and 1\,000 messages per protocol to show the impact of the trace size on the inference quality.
Fewer messages were available for AWDL and AU, which we consider in the discussion of the results.

\subsection{Pseudo Data Type Clustering Validation}\label{sec:eval-validate}

First, we validate our base assumption that data types of segments can be clustered using the Canberra dissimilarity.
\autoref{sec:step1} describes the process to cluster for pseudo data types.
For validation, we compare the clustering results to the true field data types from the Wireshark dissectors.
This provides a baseline to validate that different data types can correctly be distinguished by our dissimilarity measure.

Cluster statistics quantify the accuracy of the match between data types and clusters.
As overall quality metrics, we provide $P$, $R$, and F-score for our test protocols in \autoref{tab:gt-clusters-comb}.
For reference, we include the number of messages in the trace, the number of \emph{unique} fields in the trace, and the auto-configured $\varepsilon$.
The amount of noise identified by DBSCAN is always zero.

The F-score values in \autoref{tab:gt-clusters-comb} are near the optimum of 1 which shows that data types can be clustered with high utility.
However, the SMB trace with 1\,000 messages stays behind the other results
due to its low precision.
Inspection of the individual clusters shows that timestamps and signatures have erroneously been placed together in one cluster.
Ignoring this single cluster for the sake of the argument,
we gain a precision of 0.96 while the recall drops to 0.37.
This is the only instance in all our test runs where a parameter selection fails with such a significant impact, hinting towards great robustness of the method.
Protocols with complex message formats, like DHCP and SMB, require a large amount of variability in the trace to allow for a decent analysis result.
\autoref{tab:gt-clusters-comb} shows this by the lower F-scores and specifically the lower recall for these complex protocols with smaller traces of 100 messages compared to the results for 1\,000 messages of the same protocols.
This is due to multiple clusters representing a disjointed group of similar segments, reducing the recall.

Based on the high precision of almost all clustering results, we conclude that most field types can accurately be clustered by means of dissimilarity.
Overall, this aspect of our evaluation shows that clusters match with true field types and thus validate our approach of data type clustering.

\subsection{Clustering with Imperfect Segmentation}\label{sec:clustering-of-pseudo-data-types}

Next, we present our evaluation of clustering similar segments of real-world protocols without relying on perfect segmentation from Wireshark dissectors.
Instead, we use the existing heuristic segmenters Netzob~\cite{bossert_towards_2014}, NEMESYS~\cite{kleber_nemesys:_2018}, and CSP~\cite{goo_protocol_2019} on our set of known test protocols.
This way, we emulate the lack of ground truth during clustering while retaining the possibility to measure the inference quality.

%
We compare three existing heuristics segmenters that are available for unknown binary protocols as a basis for our field data type clustering.
According to our results, no single segmenter is clearly superior to the others and each has its strengths and weaknesses with regard to the kind of analyzed protocol.
\autoref{tab:segmenters-clusters-comb} contains the clustering statistics $P$, $R$, and the F-score per test protocol.
We mark the best-performing segmenter for each protocol trace by bold printed values in the table.
Four analysis runs fail due to exceeding runtime or memory constraints.


A significant number of segments cannot be clustered correctly and concisely as their boundaries are shifted relative to the true position they should optimally mark.
These fragments blur some segment clusters to the extent that we cannot clearly separate the affected data types.
\begin{figure}
\centering
\begin{tikzpicture}[
		node distance=0pt, yscale=2,
		every node/.style={font=\ttfamily, text height=.7em, outer sep=0, inner sep=0},
		tfe/.style={draw, minimum height=1.2em, thick},
		tfelabel/.style={rotate=-20, anchor=north west},
		ftstimestamp/.style={},
		staticmarker/.style={fill=uulm, opacity=.33}]
	
	\coordinate(m8f16) at (0,0);
	
	\node[right=0ex of m8f16, ftstimestamp] (m8f17) {d2};
	\node[right=of m8f17, ftstimestamp] (m8f18) {3d};
	\node[right=of m8f18, ftstimestamp] (m8f19) {19};
	\node[right=of m8f19, ftstimestamp] (m8f20) {03};
	\node[right=of m8f20, ftstimestamp] (m8f21) {b3};
	\node[right=of m8f21, ftstimestamp] (m8f22) {fc};
	\node[right=of m8f22, ftstimestamp] (m8f23) {da};
	\node[right=of m8f23, ftstimestamp] (m8f24) {b1};
	
	\node[right=0ex of m8f16, yshift=-3ex, ftstimestamp] (m8f25) {d2};
	\node[right=of m8f25, ftstimestamp] (m8f26) {3d};
	\node[right=of m8f26, ftstimestamp] (m8f27) {19};
	\node[right=of m8f27, ftstimestamp] (m8f28) {7a};
	\node[right=of m8f28, ftstimestamp] (m8f29) {01};
	\node[right=of m8f29, ftstimestamp] (m8f30) {58};
	\node[right=of m8f30, ftstimestamp] (m8f31) {10};
	\node[right=of m8f31, ftstimestamp] (m8f32) {62};
	
	\node[right=0ex of m8f16, yshift=-6ex, ftstimestamp] (m8f33) {d2};
	\node[right=of m8f33, ftstimestamp] (m8f34) {3d};
	\node[right=of m8f34, ftstimestamp] (m8f35) {19};
	\node[right=of m8f35, ftstimestamp] (m8f36) {1c};
	\node[right=of m8f36, ftstimestamp] (m8f37) {d0};
	\node[right=of m8f37, ftstimestamp] (m8f38) {25};
	\node[right=of m8f38, ftstimestamp] (m8f39) {d0};
	\node[right=of m8f39, ftstimestamp] (m8f40) {74};
	
	
	\node[fit=(m8f18)(m8f21), tfe] (fittedA) {};
	\node[fit=(m8f26)(m8f28), tfe] (fittedB) {};
	\node[fit=(m8f34)(m8f38), tfe] (fittedC) {};
	
	\draw[tfe] (m8f16|-fittedA.north) to (m8f24.east|-fittedA.north);
	\draw[tfe] (m8f16|-fittedA.south) to (m8f24.east|-fittedA.south);
	\draw[tfe] (m8f16|-fittedB.north) to (m8f24.east|-fittedB.north);
	\draw[tfe] (m8f16|-fittedB.south) to (m8f24.east|-fittedB.south);
	\draw[tfe] (m8f16|-fittedC.north) to (m8f24.east|-fittedC.north);
	\draw[tfe] (m8f16|-fittedC.south) to (m8f24.east|-fittedC.south);

	\node[fit=(m8f17)(m8f35), staticmarker] (static) {};

	\node[left=1em of m8f17.base west, font=\sffamily\small, anchor=base east] {NTP timestamp A};
	\node[left=1em of m8f25.base west, font=\sffamily\small, anchor=base east] {NTP timestamp B};
	\node[left=1em of m8f33.base west, font=\sffamily\small, anchor=base east] {NTP timestamp C};
\end{tikzpicture}
\caption{
Typical errors in heuristically inferred segment boundaries (vertical lines) that should approximate timestamps.
The shaded area marks static bytes.
}
\label{fig:timestamp-error}
\end{figure}
%
\autoref{fig:timestamp-error} illustrates how this affects the dissimilarity measure and, thus, the clustering result with an example of three timestamps that have incorrect additional boundaries splitting the true field.
These least significant bytes of the timestamps, regarded by themselves, seem random and thus cannot be clustered based on their value.
This error is not an effect of the dissimilarities used as segment features or the clustering algorithm, but stems from incorrect partitioning of the message by the segmenters.

This error in the approximated boundaries of high-entropy fields is the reason for SMB's low recall as it
contains a signature that is randomly split by all of the segmenters, since its contents look random across different messages.
%
AU's segments suffer from a slightly different but related issue: long sequences of 32-bit integers, representing measurement results, look static in some instances and random in others so that the dissimilarity is not successfully exploitable for clustering.
Since for AU we only have 123 messages available to evaluate, we hypothesize that the variance incurred by larger traces would have a positive impact if available.
For the other traces of different sizes of all protocols, the precision stays high compared to the true-fields baseline (\autoref{sec:eval-validate}).

Considering the best case per protocol, only the larger trace of SMB exhibits an unsatisfactory precision of of 0.57, which is still remarkable, since knowing the true segments leads to only a very small improvement ($R=0.59$).
The smaller SMB trace and the AU trace are unsatisfying due to their low recall while precision in both cases remains high.
We marked the three unsatisfying cases by red-colored F-scores and in contrast colored all F-scores of at least 0.8 green, which we consider successful analyzes.
Most of the results even score above 0.9 with a precision of also better than 0.9.
In the face of the identified problems that are realistic for working with unknown protocols, we argue that our method can cope with the inaccurate segmentation to a large degree.

The remaining challenge is to select the most suited segmenter for a protocol trace.
We see that Netzob is most suited for protocols with distinct patterns of repeating value sequences\eg NTP having fixed structure and AWDL with a type-length-value (TLV) record structure.
Large messages cause Netzob to fail due to the exponential increase in runtime, which is the case for larger traces of DHCP and SMB, and for the AU trace.
NEMESYS deals well with large and complex messages, especially since they contain a mixture of number values and chars, which fits the heuristic of NEMESYS best.
CSP performs minimally worse for larger traces than NEMEYS, but it lags behind for smaller traces.
As CSP is more dependent an the variance in the trace, it is best applied to large traces where it poses an alternative to NEMESYS.

\subsection{Evaluation Summary}

This evaluation provides two insights about our approach:
(1) field data type clustering works as intended with very little requirements towards and assumptions about the protocols, but
(2) field data type clustering highly depends on the segmentation result,
where we rely on existing approaches that provide results of only limited quality.
The higher the correctness of the heuristic segmentation, the better the message field type clustering can perform.

In comparison, FieldHunter is able to discern the concrete data type of typically one or two fields per message, leading to a coverage of 3\,\% on average across all protocols. 
While, in contrast, our clustering method per se cannot determine the field type, it achieves an average coverage of 87\,\% (see \autoref{tab:segmenters-clusters-comb}), which means that we can provide information about the structure of messages in terms of field similarity and field's value domains for almost the complete content of all messages.

\section{Conclusion}
\label{sec:conclusion}

In this paper, we propose a novel method to cluster field data types in messages of unknown binary protocols.
It requires recorded network traces and leverages the similarity of segments to group them into clusters representing a common data type.
Our efficient clustering of message segments facilitates subsequent analyses to identify their likely semantic function.
We envision that identified data types and visual analytics will
improve the analysis efficiency of unknown network messages by providing the means to determine the most security relevant message parts to investigate further in a given trace.

In PRE, a typical high-effort task is to understand the large-scale structure of messages.
Knowing such structure is often the basis to analyze\eg data exfiltration by malware, privacy violations, targets for spoofing and fuzzing for vulnerability testing as we illustrated\eg in \citet{stute_billion_2019,kroll_aristoteles_2021}.
Automating this process saves effort and time and our work contributes to such automation.
Opposed to previous work that uses a set of heuristics to recognize a fixed number of field types, clustering of segments is also applicable if the protocol contains unanticipated data representations\eg encodings, since it only relies on the segments' similarity and occurrence.
Thus, we can cover large parts of the messages in the trace, while previous work---with a coverage of only 3\,\% on average---leaves most of the message content completely unintelligible.
While clustering per se does not reveal data types, it simplifies an analyst's interpretation of the message content.
Our method increases the coverage of the interpretable message content to 87\,\% on average, outperforming the state-of-the-art by almost factor 30 and enabling comprehension of the large-scale message structure.

We first evaluated our approach for both publicly documented as well as undocumented protocols relying on ground truth message fields derived from Wireshark dissectors.
We find that most field data types can be clustered with high precision when knowing correct field boundaries.
In realistic situations, where field boundaries are not known and heuristic segmenters like Netzob, NEMESYS, or CSP are applied, the recall is lower, but data types can still be distinguished with a precision close to 100\,\% in most cases.
Our approach works also for protocols without IP encapsulation, like AWDL and AU, where previous work could not be applied due to the field type heuristics' reliance on context information.

We see two main areas for future work.
Firstly, we propose to combine our data type clustering with the deduction of intra- and inter-message semantics similar to FieldHunter~\cite{bermudez_towards_2016}.
This would enable the interpretation of\eg length fields and message counter fields.
Moreover, we intend to automatically learn value generation rules from the cluster contents 
using LSTM or similar machine learning methods 
to predict probable field values for fuzzing and misbehavior detection.
\section*{Acknowledgment}
We would like to thank Steffen Klee for providing us with a Wireshark dissector and traces for Apple's Auto Unlock (AU) protocol.

\printbibliography

%

\end{document}